\documentstyle[12pt,A4]{article}
\begin{document}
\label{zacatek-vys4}
\noindent
\hspace*{8cm}{\sc acta univ. palacki. olomuc.,}\\
\hspace*{8cm}{\sc fac.\,rer.\,nat.\,{\small (1998)},\,physica\,{\small
37}},\\
\hspace*{8cm}{\small \pageref{zacatek-vys4} -- \pageref{konec-vys4}}\\
\hspace*{8cm}\rule[3mm]{6.2cm}{0.2mm}\\[1cm]
\centerline{\large \bf THE REVISED QUANTUM MECHANICAL THEORY}\\
\centerline{\large \bf OF THE OPTICAL ACTIVITY OF CRYSTALS}\\[2mm]
\centerline{\bf Ivo Vy\v s\'\i n, Jan \v R\'\i ha} \\[3mm]
\noindent {\small Department of Theoretical Physics, \ Natural Science
Faculty of Palack\' y University, \ Svobody 26, {\bf 771~46 Olomouc}, \
Czech Republic}\\[4mm]
\centerline{\it Received 26th November 1997}
\vspace*{0.5cm}

{\sc ABSTRACT:} \ In this paper we present the revised view on
the optical activity of
crystals based on the model of two dumped coupled oscillators. The
results are compared with the results of the same problem solved before
and it is presented that the results of the quantum mechanical model are the
generalization of the classical model only and they do not introduce the new
terms of quantum mechanical nature. The results are discussed with regard to
the rotational strengths of normal modes of vibrations and the formula
for the complex rotatory power $\bar \rho (\omega )$ containing the
rotational strengths is presented.

\section{Introduction}
\hspace \parindent
Optical activity (OA) is manifested by dissymmetric species because the
transition probabilities are different for the left and the right circularly
polarized wave. This means a difference between the complex refractive
indices that we denote $\bar n_{l}$ for the left and $\bar n_{r}$ for the
right circularly polarized wave.
The complex rotatory power may be considered
to be complex quantity given by
\begin{equation}
\label{1}
\bar \rho =\rho +i\sigma =\frac{\omega }{2c}\left (\bar n_{l}-\bar
n_{r}\right )=\frac{\omega }{2c}\left [\left (n_{l}-n_{r}\right )+i\left
(\kappa _{l}-\kappa _{r}\right )\right ],
\end{equation}
where $\rho $ is the rotation angle of incidenting linear polarized
light per unit length,
and for a very excellent approximation $\sigma $ is the ellipticity per
unit length. The variation of $\rho $ with frequency $\omega $ or
wavelength $\lambda $ is called the optical rotatory dispersion (ORD) and
the variation of $\sigma $ is the circular dichroism (CD); $n_{l}$, $n_{r}$ and
$\kappa _{l}$, $\kappa _{r}$ are real and imaginary parts of the complex
refraction indices $\bar n_{l}$ and $\bar n_{r}$. So that the OA has both
aspects arising from interaction of radiation with matter - dispersive
and absorptive and these aspects are connected by the Kramers - Kronig
relations.

The important group of the optically active crystals are the crystals
with screw axis of symmetry belonging to the space group of symmetry
$D_{3}^{4}$ and its enantiomorphic $D_{3}^{6}$. The typical
representatives of these crystals are $\alpha $-quartz, cinnabar,
tellurium, selen, camphor and benzil. The optical activity of these
crystals is caused by the assymetrical originating of the crystal structure
because the molecules or atoms forming the crystals are symmetrical.
Camphor is the
exception because its molecules are optically active.
It means that the other crystals belonging to these groups of symmetry
are optically active in the crystalline state only.

The dispersion of the optical activity was studied in works based on the
excitons theory \cite{Agr1,Agr2,Tsv,Kato}, on the theory of coupled
oscillators \cite{Chand1,Chand2,Vys1,Vys2,Janku} or on the Lagrangian
formalism \cite{Nelson}. But all these models based on the exciton theory,
some models of coupled oscillators and the model in \cite{Nelson} are
solved in the frequency region far from the absorption range. It is well known
that the CD is nozeroth only in a very narrow
frequency region in the absorption range and therefore these models solve
only ORD as one part of the optical activity independently on the CD.

Both aspects of optical activity can be solved by means of the model of
the dumped coupled oscillators. Vy\v s\'\i n \cite{Vys2} has solved this
model semiclassically and Jank\accent23 u \cite{Janku} by quantum
mechanical way where the role of dumping plays the limited lifetime of
oscillators in excited
states. But Jank\accent23 u has got some other terms in comparison with
the results of \cite{Vys2} in his formulae for ORD and CD
and he supposes that these terms are of quantum mechanical nature.
At the same time the sense of these terms is not evident. For this
reason we revise the solution of this problem.

All above mentioned works that solve the optical activity of crystals
by the coupled
oscillator model arised from the Kuhn model \cite{Kuhn} of two
coupled oscillators forming the compound oscillator. It is well known
that two asymmetrically oriented harmonic oscillators coupled
together form the fundamental optically active unit. With regard
to the structure of crystals belonging to the space groups of symmetry
$D_{3}^{4}$ and $D_{3}^{6}$ we can use the following Chandrasekhar
model of compound oscillator.

We assume that the coordinate axis $z$ is parallel with the crystal
axis $c$.
The first single dumped linear harmonic oscillator lies in the plane
$z=0$ and his direction of vibrations is given by direction cosines
$\alpha $, $\beta $, $\gamma $. The second oscillator lies in the plane
$z=d$ where as $d$ we denote the projection of distance between
oscillators into the $z$ axis and his direction of vibrations is turned
by the angle $\theta $ around axis $z$ with respect to the first one. The
direction of vibrations of the second oscillator is so given by the
cosines $\alpha \cos \theta -\beta \sin \theta$, $\alpha \sin \theta
+\beta \cos \theta $, $\gamma $. Both oscillators lie on the helix
which is given by the structure of crystal. The angle $\theta $ is, of
course, $120$ deg for
the crystals belonging to the space groups of symmetry $D_{3}^{4}$ and
$D_{3}^{6}$.

Both oscillators forming the compound oscillator are identical. They
have the identical mass, electric charge and lifetime in excited states.
We take the
interaction between adjacent oscillators as week, that is for
example as dipol - dipol character. We suppose that the left and
the right circularly polarized wave, into which
the incidenting linear polarized wave is decomposed in
the optically active crystal, propagate only along the
crystal axis. This is the most practically important case because the
optical activity is not masked by birefringence.

\section{The quantum mechanical solution of the model}
\hspace \parindent
We can set the Schr\H odinger equation for the one compound
oscillator in the field of the left and the right circularly polarized wave
\begin{equation}
\label{2}
\hat {\cal H}\Psi=-\frac{\hbar ^{2}}{2m}\sum_{\xi =1}^{2}\frac{\partial
^{2}\Psi }{\partial r_{\xi}^{2}}+\frac{m\omega _{0}^{2}}{2}\sum_{\xi
=1}^{2}r_{\xi }^{2}\Psi +\mu r_{1}r_{2}\Psi+\frac{i}{m\omega }\sum_{\xi
=1}^{2}F_{\xi }^{l,r}\hat p_{\xi}e^{\gamma _{0} t}\Psi ,
\end{equation}
where $r_{1}$, $r_{2}$ are the displacements of oscillators from
equilibrium, $\mu r_{1}r_{2}$ is the potential energy of mutual
interactions of the oscillators, $F_{1}^{l,r}=eE_{1}^{l,r}$,
$F_{2}^{l,r}=eE_{2}^{l,r}$ are the
electric forces projections of the left and the right circularly polarized
wave into the vibration directions of the oscillators, $e$ is the
charge of electron. The upper
index $l,r$ holds for the left and the right circularly polarized wave,
${\hat p}_{1}$ and ${\hat p}_{2}$ are the moment operators of both
oscillators. The
small positive parameter $\gamma _{0} $ gives the possibility of the adiabatic
interaction at the time $t=-\infty $.
The dumping due to the limited lifetime of oscillators in their
excited states is formally introduced by the parameter $\gamma _{0}$.

The electric vector $\vec E$ of the left and the right circularly polarized
wave propagating along $z$ axis has the components
\begin{eqnarray}
\label{3}
E_{x}^{l}=E_{0}e^{-i\left (\omega t-k_{l}z\right )},&\qquad &
E_{y}^{l}=E_{0}e^{-i\left (\omega t-k_{l}z-\frac{\pi }{2}\right )}
\nonumber \\
E_{x}^{r}=E_{0}e^{-i\left (\omega t-k_{r}z\right )},&\qquad &
E_{y}^{r}=E_{0}e^{-i\left (\omega t-k_{r}z+\frac{\pi }{2}\right )}
\end{eqnarray}
and the vector $\vec E$ projections of these waves into the directions of
vibrations of oscillators are
\begin{eqnarray}
\label{4}
E_{1}^{l,r}&=&E_{0}e^{-i\omega t}\left (\alpha +\beta e^{\pm i\frac{\pi
}{2}}\right ) \nonumber \\
E_{2}^{l,r}&=&E_{0}e^{-i\omega t}\left[ \left (\alpha \cos \theta -\beta
\sin \theta \right )e^{i\phi _{l,r}}+\left (\alpha \sin \theta +\beta
\cos \theta \right )e^{i\left (\phi _{l,r}\pm \frac{\pi }{2}\right
)}\right],
\end{eqnarray}
where $\phi _{l,r}=k_{l,r}={\bar n}_{l,r}\omega d/c$ is a complex phase
shift and
${\bar n}_{l,r}$ are refractive indices of medium for the left (index $l$) and
the right (index $r$) circularly polarized wave. In all terms in eq.
(\ref{4}) and further the $+$ sign holds for the left and
the $-$ sign for the right circularly polarized wave.

Due to the mutual interaction between coupled oscillators the natural
frequency $\omega _{0}$ splits into two adjacent frequencies of the normal
modes of vibrations. Introducing the normal coordinates $q_{1}$ and
$q_{2}$ which are given by the relations
\begin{equation}
\label{5}
r_{1}=\frac{1}{\sqrt{2}}\left (q_{1}+q_{2}\right ),\qquad
r_{2}=\frac{1}{\sqrt{2}}\left (q_{1}-q_{2}\right ),
\end{equation}
into eq. (\ref{2}) we get
\begin{eqnarray}
\label{6}
\hat {\cal H}\Psi &=&-\frac{\hbar ^{2}}{2m}\sum_{\eta
=1}^{2}\frac{\partial ^{2}\Psi }{\partial q_{\eta}^{2}}+\frac{m\omega
_{0}^{2}}{2}\sum_{\eta =1}^{2}q_{\eta }^{2}\Psi +\frac{1}{2}\mu \left
(q_{1}^{2}-q_{2}^{2}\right )\Psi \nonumber \\
&&+\frac{i}{m\omega }\left[ \frac{1}{\sqrt {2}}\left
(F_{1}^{l,r}+F_{2}^{l,r}\right )\hat p_{q_{1}}+\frac{1}{\sqrt {2}}\left
(F_{1}^{l,r}-F_{2}^{l,r}\right )\hat p_{q_{2}}\right] e^{\gamma
_{0}t}\Psi .
\end{eqnarray}

It may be easily verified that in the eq. (\ref{6}) the expresions
$\frac{1}{\sqrt{2}}\left (F_{1}^{l,r}+F_{2}^{l,r}\right )$ and
$\frac{1}{\sqrt{2}}\left (F_{1}^{l,r}-F_{2}^{l,r}\right )$ are the
projections of electric forces in normal coordinates. We denote them as
$F_{q_{1}}^{l,r}$ and $F_{q_{2}}^{l,r}$.

Now the equation (\ref{6}) can be separated by means of
\begin{eqnarray}
\label{7}
\Psi (q_{1},q_{2},t)&=&\Psi _{1}(q_{1},t)\Psi _{2}(q_{2},t),\nonumber \\
\hat {\cal H}&=&\hat {\cal H}_{q_{1}}+\hat {\cal H}_{q_{2}}.
\end{eqnarray}

Substituting (\ref{7}) into (\ref{6}), further dividing by $\Psi
_{1}(q_{1},t)\Psi _{2}(q_{2},t)$ and posing the member with $q_{1}$ to
$\hat {\cal H}_{q_{1}}$ and $q_{2}$ to $\hat {\cal H}_{q_{2}}$ we get
two equations
\begin{eqnarray}
\label{8}
\hat {\cal H}_{q_{1}}\Psi _{1}(q_{1},t)&=&-\frac{\hbar
^{2}}{2m}\frac{\partial ^{2}\Psi _{1}(q_{1},t)}{\partial
q_{1}^{2}}+\frac{m\omega _{0}^{2}+\mu }{2}q_{1}^{2}\Psi
_{1}(q_{1},t)\nonumber \\
&&+\frac{i}{m\omega }\left (F_{q_{1}}^{l,r}\hat p_{q_{1}}\right
)e^{\gamma _{0}t}\Psi _{1}(q_{1},t)\nonumber \\
\hat {\cal H}_{q_{2}}\Psi _{2}(q_{2},t)&=&-\frac{\hbar
^{2}}{2m}\frac{\partial ^{2}\Psi _{2}(q_{2},t)}{\partial
q_{2}^{2}}+\frac{m\omega _{0}^{2}-\mu }{2}q_{2}^{2}\Psi
_{2}(q_{2},t)\nonumber \\
&&+\frac{i}{m\omega }\left (F_{q_{2}}^{l,r}\hat p_{q_{2}}\right
)e^{\gamma _{0}t}\Psi _{2}(q_{2},t)
\end{eqnarray}
and we see that the natural frequency is split into two vibrations of
the normal modes. For these frequencies we have
\begin{equation}
\label{9}
\omega _{1}^{2}=\omega _{0}^{2}+Q,\qquad \omega _{2}^{2}=\omega
_{0}^{2}-Q;
\end{equation}
$Q=\mu /m$.

For $F_{q_{1}}^{l,r}$ and $F_{q_{2}}^{l,r}$ in the equations (\ref{8})
we can derive using eq. (\ref{4}) the expression
\begin{equation}
\label{10}
F_{q_{\eta }}^{l,r}=e\left (a_{q_{\eta }}^{l,r}\right )E_{0}e^{-i\left (\omega t+\sigma
_{q_{\eta }}^{l,r}\right )};\quad \eta =1,2.
\end{equation}
The coefficients $a_{q_{\eta}}^{l,r}$ are determined by the relations
\begin{eqnarray}
\label{11}
\left (a_{q_{1}}^{l,r}\right )^{2}&=&\left (\alpha ^{2}+\beta
^{2}\right )\left (1+\cos \theta \mp \phi _{l,r}\sin \theta \right
),\nonumber \\
\left (a_{q_{2}}^{l,r}\right )^{2}&=&\left (\alpha ^{2}+\beta
^{2}\right )\left (1-\cos \theta \pm \phi _{l,r}\sin \theta \right );
\end{eqnarray}
$\sigma _{q_{\eta }}^{l,r}$ are the meaning of the phase shifts only. In
the eqs. (\ref{11}) the upper sign holds
for the left and the lower sign for the right circularly polarized wave. The
eqs. (\ref{8}) can be rewritten in the general form
\begin{eqnarray}
\label{12}
\lefteqn{\hat {\cal H}_{q_{\eta }}\Psi _{\eta }(q_{\eta },t)=-\frac{\hbar
^{2}}{2m}\frac{\partial ^{2}\Psi _{\eta }(q_{\eta },t)}{\partial
q_{\eta }^{2}}+\frac{m\omega _{\eta }^{2}}{2}q_{\eta }^{2}\Psi
_{\eta }(q_{\eta },t)}\nonumber \\
&&+\frac{ie}{m\omega }\left (a_{q_{\eta }}^{l,r}\right )E_{0}
\hat p_{q_{\eta }}e^{-i\left (\omega t+\sigma _{q_{\eta }}^{l,r}\right )
+\gamma _{0}t}\Psi _{\eta }(q_{\eta },t);
\end{eqnarray}
$\eta =1,2$. The Schr\H odinger equations for the normal modes of
vibrations then are
\begin{equation}
\label{13}
i\hbar \frac{\partial \Psi _{\eta }(q_{\eta},t)}{\partial t}=\hat {\cal
H}_{q_{\eta }}^{{l,r}^{0}}\Psi _{\eta }(q_{\eta },t)+\hat {\cal
H}_{q_{\eta }}^{{l,r}^{p}}\Psi _{\eta }(q_{\eta },t),
\end{equation}
where $\hat {\cal H}_{q_{\eta }}^{{l,r}^{0}}$ is a nonperturbed and
$\hat {\cal H}_{q_{\eta }}^{{l,r}^{p}}$ a perturbed hamiltonian. In our
case
\begin{equation}
\label{14}
\hat {\cal H}_{q_{\eta }}^{{l,r}^{0}}=-\frac{\hbar
^{2}}{2m}\frac{\partial ^{2}}{\partial q_{\eta }^{2}}+\frac{m\omega
_{\eta }^{2}}{2}q_{\eta }^{2}
\end{equation}
and we see that the nonperturbed hamiltonian depends only on the index $\eta
$ of
the normal mode of vibrations and it doesn't depend
on the polarization of the light
wave. We can further write $\hat {\cal H}_{q_{\eta }}^{{l,r}^{0}}=\hat
{\cal H}_{q_{\eta }}^{0}$.
On the other hand the perturbed hamiltonian
\begin{equation}
\label{15}
\hat {\cal H}_{q_{\eta }}^{{l,r}^{p}}=\frac{ie}{m\omega }\left
(a_{q_{\eta }}^{l,r}\right )E_{0}{\hat p}_{q_{\eta }}e^{-i\left (\omega
t+\sigma _{q_{\eta }}^{l,r}\right )+\gamma _{0}t}
\end{equation}
depends on the index of the mode and also on the polarization of the wave.

The mean value of the induced electric dipol moment from the side of the left
and the right circularly polarized wave that we hold as small perturbation we can
solve by means of the Kubo theorem \cite{Davyd}
\begin{equation}
\label{16}
\left \langle \overline {e\left(a_{q_{\eta
}}^{l,r}\right)q_{\eta }}\right \rangle =\left
\langle e\left (a_{q_{\eta }}^{l,r}\right )q_{\eta }\right \rangle
_{\eta _{0}}+\frac{ie^{2}}{m\omega }\left(a_{q_{\eta }}^{l,r}\right)^{2}E_{0}\left
\langle \left \langle {\hat q}_{\eta },{\hat p}_{q_{\eta }}\right
\rangle \right \rangle _{\omega }e^{-i\left (\omega
t+\sigma _{q_{\eta }}^{l,r}\right )+\gamma _{0}t}.
\end{equation}

The first term on the right side of eq. (\ref{16}) is the constant
dipole moment of system. This term has no
meaning for us. The expression $\left \langle \left \langle {\hat q}_{\eta
},{\hat p}_{q_{\eta }}\right \rangle \right \rangle _{\omega }$
is the Fourier transform of the retarded Green function $\left \langle \left
\langle {\hat q}_{\eta },{\hat p}_{q_{\eta }}\right \rangle \right
\rangle _{t}$ of operators
${\hat q}_{\eta }$ and ${\hat p}_{q_{\eta }}$, that is
\begin{equation}
\label{17}
\left \langle \left \langle {\hat q}_{\eta },{\hat p}_{q_{\eta
}}\right \rangle \right \rangle _{\omega }=\frac{1}{\hbar }\int
\limits_{-\infty }^{\infty }e^{i\omega
t-\gamma _{0}t}\left \langle \left
\langle {\hat q}_{\eta },{\hat p}_{q_{\eta
}}\right \rangle \right \rangle _{t}dt.
\end{equation}

For the retarded Green function $\left \langle \left
\langle {\hat q}_{\eta },{\hat p}_{q_{\eta }}\right \rangle \right
\rangle _{t}$ in the ground states of quantum system $\vert \eta _{0}\rangle $
we hold the relation
\begin{equation}
\label{18}
\left \langle \left \langle {\hat q}_{\eta },{\hat p}_{q_{\eta }}\right
\rangle \right \rangle _{t}=-i\vartheta (t)\langle \eta _{0}\vert
\left[{\hat{\tilde q}}_{\eta }(t),{\hat p}_{q_{\eta }}\right]\vert
\eta _{0}\rangle .
\end{equation}

In the eq. (\ref{18}) $\vartheta (t)$ is the unit step function,
${\hat{\tilde q}}_{\eta }(t)$ is the operator of $q_{\eta }$ in the
interactions representation
\begin{equation}
\label{19}
{\hat{\tilde q}}_{\eta }(t)=e^{i{\hat {\cal H}}_{q_{\eta }}^{0}t/\hbar
}{\hat q}_{\eta }e^{-i{\hat {\cal H}}_{q_{\eta }}^{0}t/\hbar }
\end{equation}
and for the aplication of the operator in the exponent on the wave function
holds
\begin{equation}
\label{20}
e^{\pm i{\hat {\cal H}}_{q_{\eta }}^{0}t/\hbar }\vert \eta _{n}\rangle
=e^{\pm iE_{\eta _{n}}t/\hbar }\vert \eta _{n}\rangle =e^{\pm i\omega _{\eta
_{n}}t}\vert \eta _{n}\rangle .
\end{equation}

In following rearangement we use that for matrix elements of the production
of operators ${\hat F}_{q_{\eta }}$ and ${\hat K}_{q_{\eta }}$ holds in
the normal modes the relation
\begin{equation}
\label{21}
\langle \eta _{m}\vert \hat F_{q_{\eta }}\hat K_{q_{\eta }}\vert \eta
_{n}\rangle =\sum_{k}\langle \eta _{m}\vert \hat
F_{q_{\eta }}\vert \eta _{k}\rangle \langle \eta _{k}\vert \hat K_{q_{\eta
}}\vert \eta _{n}\rangle
\end{equation}
and further we use that the matrix elements of operator ${\hat p}_{q_{\eta
}}$ are
\begin{equation}
\label{22}
\langle \eta _{m}\vert {\hat p}_{q_{\eta }}\vert \eta _{n}\rangle =im\omega
_{\eta _{mn}}\langle \eta _{m}\vert {\hat q}_{\eta }\vert \eta
_{n}\rangle ;
\end{equation}
$\omega _{\eta _{mn}}=\omega _{\eta _{m}}-\omega _{\eta _{n}}$.

Then we can solve that
\begin{equation}
\label{23}
\langle \eta _{0}\vert
\left[{\hat{\tilde q}}_{\eta }(t),{\hat p}_{q_{\eta }}\right]\vert
\eta _{0}\rangle =im\sum_{k}\omega _{\eta _{k0}}\vert \langle \eta
_{k}\vert {\hat q}_{\eta }\vert \eta _{0}\rangle \vert
^{2}\cdot \left(e^{-i\omega _{\eta _{k0}}t}+e^{i\omega _{\eta _{k0}}t}\right)
\end{equation}
and after substituting the result of eq. (\ref{23}) into (\ref{18}) and
solving the integral on the right side of eq. (\ref{17}) we have
\begin{equation}
\label{24}
\left \langle \left \langle {\hat q}_{\eta },{\hat p}_{q_{\eta }}\right
\rangle \right \rangle _{\omega }= \frac{2i\omega m}{\hbar
}\sum_{k}\frac{\omega _{\eta _{k0}}\vert \langle \eta _{k}\vert {\hat q}_{\eta
}\vert \eta _{0}\rangle \vert ^{2}}
{\omega ^{2}-\omega _{\eta _{k0}}^{2}+2i\gamma _{0}\omega }.
\end{equation}

The induced dipole moments $d_{q_{\eta }}^{l,r}$ we can solve by means
of (\ref{24}) and (\ref{16})
\begin{eqnarray}
\label{25}
d_{q_{\eta }}^{l,r}&=&\sum_{k}\frac{2e^{2}
\left(a_{q_{\eta }}^{l,r}\right)^{2}\omega _{\eta _{k0}}\vert \langle
\eta _{k}\vert {\hat
q}_{\eta }\vert \eta _{0}\rangle \vert ^{2}}{\hbar \left(\omega
_{\eta _{k0}}^{2}-\omega ^{2}-2i\gamma _{0}\omega \right)}\nonumber \\
&&\times E_{0}e^{-i\left (\omega
t+\sigma _{q_{\eta }}^{l,r}\right )+\gamma _{0}t}
\end{eqnarray}
and we can introduce the oscillator strengths of
the normal modes of vibrations into (\ref{25}) by the relation
\begin{equation}
\label{26}
f_{q_{\eta }}=\frac{2m\omega _{\eta _{k0}}\vert \langle \eta _{k}\vert
q_{\eta }\vert \eta _{0}\rangle \vert ^{2}}{\hbar }
\end{equation}
Now the induced dipole moments can be expressed as
\begin{equation}
\label{27}
d_{q_{\eta }}^{l,r}=\sum_{k}\left(a_{q_{\eta
}}^{l,r}\right)^{2}\frac{e^{2}f_{q_{\eta }}}{m}\frac{E_{0}e^{-i\left (\omega
t+\sigma _{q_{\eta }}^{l,r}\right )+\gamma _{0}t}}{\omega
_{\eta _{k0}}^{2}-\omega ^{2}-2i\gamma _{0}\omega }.
\end{equation}

The mean polarizability per unit volume would be then
\begin{equation}
\label{28}
\chi _{q_{\eta }}^{l,r}=\frac{Nd_{q_{\eta }^{l,r}}}{E_{0}e^{-i\left (\omega
t+\sigma _{q_{\eta }}^{l,r}\right )+\gamma _{0}t}},
\end{equation}
where $N$ is the number of coupled oscillators in a volume unit. But if we
include all coupling in the crystal then we see that the number of
coupling between two adjacent oscilators in the direction of propagating
light is the same as the number of single oscillators.

Substituting from (\ref{28}) and (\ref{27}) into the Drude - Sellmaier
dispersion relation of refractive indices of the crystals
\begin{equation}
\label{29}
{\bar n}_{l,r}^{2}-1=4\pi \sum_{\eta =1}^{2}\chi _{q_{\eta }}^{l,r}
\end{equation}
we have
\begin{equation}
\label{30}
{\bar n}_{l,r}^{2}-1=\frac{4\pi Ne^{2}}{m}\sum_{\eta
=1}^{2}\sum_{k}\left(a_{q_{\eta }}^{l,r}\right)^{2}\frac{f_{q_{\eta }}}{\omega
_{\eta _{k0}}^{2}-\omega ^{2}-2i\gamma _{0}\omega }
\end{equation}
and we are able to solve the relation
\begin{equation}
\label{31}
{\bar n}_{l}^{2}-{\bar n}_{r}^{2}=\frac{4\pi Ne^{2}}{m}\sum_{\eta
=1}^{2}\sum_{k}\frac{\left[\left(a_{q_{\eta
}}^{l}\right)^{2}-\left(a_{q_{\eta }}^{r}\right)^{2}\right]f_{q_{\eta
}}}{\omega _{\eta _{k0}}^{2}-\omega ^{2}-2i\gamma _{0}\omega }
\end{equation}
that after substituting from (\ref{11}) gives
\begin{eqnarray}
\label{32}
\lefteqn{{\bar n}_{l}^{2}-{\bar n}_{r}^{2}=\frac{4\pi Ne^{2}}{m}(\alpha ^{2}+\beta ^{2})\sin
\theta \left(\phi _{l}+\phi _{r}\right)}\nonumber \\
&&\times \sum_{k}\left[\frac{-f_{q_{1}}}{\omega _{1_{k0}}^{2}-\omega
^{2}-2i\gamma _{0}\omega }+\frac{f_{q_{2}}}{\omega _{2_{k0}}^{2}-\omega
^{2}-2i\gamma _{0}\omega }\right].
\end{eqnarray}

We know that for the complex rotatory power we have the relation
${\bar \rho}(\omega )=\frac{\omega }{2c}\left({\bar n}_{l}-{\bar
n}_{r}\right)$, further $\phi _{l}+\phi _{r}=\omega d\left({\bar
n}_{l}+{\bar n}_{r}\right)/c$ and ${\bar n}_{l}^{2}-{\bar
n}_{r}^{2}=\left({\bar n}_{l}+{\bar n}_{r}\right)\left({\bar
n}_{l}-{\bar n}_{r}\right)$. Using these relation we get for
${\bar \rho }\left(\omega \right)$ the formula
\begin{eqnarray}
\label{33}
\lefteqn{\bar \rho \left(\omega \right)=\frac{2\pi Nde^{2}}{mc^{2}}\omega
^{2}\left(\alpha ^{2}+\beta ^{2}\right)\sin \theta} \nonumber \\
&&\times \sum_{k}\left[\frac{-f_{q_{1}}}{\omega _{1_{k0}}^{2}-\omega
^{2}-2i\gamma _{0}\omega }+\frac{f_{q_{2}}}{\omega _{2_{k0}}^{2}-\omega
^{2}-2i\gamma _{0}\omega }\right].
\end{eqnarray}

Now we may find the real and the imaginary part of ${\bar \rho }(\omega
)$ that from (\ref{1}) are ORD and CD
\begin{equation}
\label{34}
\mbox{Re}{\bar \rho }(\omega )=\rho (\omega )=A\omega ^{2}
\cdot \sum_{k}\left[\frac{-f_{q_{1}}\left(\omega _{1_{k0}}^{2}-\omega
^{2}\right)}{\left(\omega _{1_{k0}}^{2}-\omega ^{2}\right)^{2}+4\gamma
_{0}^{2}\omega ^{2}}+\frac{f_{q_{2}}\left(\omega _{2_{k0}}^{2}-\omega
^{2}\right)}{\left(\omega _{2_{k0}}^{2}-\omega ^{2}\right)^{2}+4\gamma
_{0}^{2}\omega ^{2}}\right],\nonumber \\
\end{equation}
\begin{equation}
\label{35}
\mbox{Im}{\bar \rho }(\omega )=\sigma (\omega
)=2A\gamma_{0}\omega ^{3}
\cdot \sum_{k}\left[\frac{-f_{q_{1}}
}{\left(\omega _{1_{k0}}^{2}-\omega ^{2}\right)^{2}+4\gamma
_{0}^{2}\omega ^{2}}+\frac{f_{q_{2}}
}{\left(\omega _{2_{k0}}^{2}-\omega ^{2}\right)^{2}+4\gamma
_{0}^{2}\omega ^{2}}\right],\nonumber \\
\end{equation}
where $A=\frac{2\pi Nde^{2}}{mc^{2}}(\alpha ^{2}+\beta ^{2})\sin \theta
$.

We know that the quantum transition is optically active when it
has a nonzeroth value of the rotational strength. It is well known that the
formulae for ORD and CD are often expressed as the sum of products of
rotational strengths for different transitions and frequency dependent
factors \cite{Deutsche}. Our formulae (\ref{34}) and (\ref{35}) for ORD
and CD do not contain the rotational strengths explicitly. But we
have showed in the paper \cite{Vysr} that we can express our formulae in
terms of rotational strengths in a similar way.

For a transition from some ground state $\vert 0\rangle $ to the excited
state $\vert k\rangle $ the rotational strength is given by imaginary
part of the scalar products of electric and magnetic dipole moments $\langle
0\vert \hat {\vec d}\vert k\rangle $ and $\langle k\vert \hat {\vec
m}\vert 0\rangle $. The rotational strength is so
\begin{equation}
\label{36}
R_{0k}={\mbox Im}\left(\langle
0\vert \hat {\vec d}\vert k\rangle \langle k\vert \hat {\vec
m}\vert 0\rangle \right).
\end{equation}

In our case of coupled oscillators the ground and the excited state
are split into two states $\vert \eta _{0}\rangle $ and $\vert \eta
_{k}\rangle $, $\eta =1,2$. Therefore the rotational strengths must be
considered between corresponding ground and excited states and they are
\begin{equation}
\label{36a}
R_{\eta _{0k}}={\mbox Im}\left(\langle \eta _{0}\vert {\hat {\vec
d}}_{q_{\eta }}\vert \eta _{k}\rangle \langle \eta _{k}\vert {\hat {\vec
m}}_{q_{\eta }}\vert \eta _{0}\rangle \right);\quad \eta =1,2
\end{equation}
and for these rotational strengths we have derived in \cite{Vysr} the
results
\begin{equation}
\label{48}
R_{1_{0k}}=-\frac{\hbar e^{2}d(\alpha ^{2}+\beta ^{2})f_{q_{1}}\sin
\theta }{8mc}
\end{equation}
and
\begin{equation}
\label{49}
R_{2_{0k}}=\frac{\hbar e^{2}d(\alpha ^{2}+\beta ^{2})f_{q_{2}}\sin
\theta }{8mc}.
\end{equation}

We see that all terms of those rotational strengths that depends
on the crystal structure are comprehended in the formula for the complex
rotatory power (\ref{33}) and consequentially in the formulae (\ref{34})
and (\ref{35}) for ORD and CD.

\section{Discussion}
\hspace \parindent
In the relations (\ref{34}) and (\ref{35}) we restrict only to
one optically active quantum
mechanical transition from the ground states $\vert \eta _{0}\rangle $
to the first
excited states $\vert \eta _{1}\rangle $. In this case we have for the transition
frequencies the relations $\omega _{1_{10}}^{2}=\omega _{1}^{2}=\omega
_{0}^{2}+Q$ and $\omega _{2_{10}}^{2}=\omega _{2}^{2}=\omega
_{0}^{2}-Q$.

The formulae (\ref{34}) and (\ref{35}) can be rewritten in the following
way. The results for two-oscillator model of optical activity are often
expressed as the sum of two terms. The first term contains the
difference of oscillators strengths $f_{q_{2}}-f_{q_{1}}$ and the second
one the sum $f_{q_{1}}+f_{q_{2}}$. If we neglect the members containing
$Q^{2}$ and $Q\gamma _{0}^{2}$ we can write for ORD and CD the relations
\begin{equation}
\label{50}
\rho (\omega )=A\omega
^{2}\Bigg\{\frac{\left(f_{q_{2}}-f_{q_{1}}\right)\left(\omega
_{0}^{2}-\omega ^{2}\right)}{\left(\omega _{0}^{2}-\omega
^{2}\right)^{2}+4\gamma _{0}^{2}\omega ^{2}}
+\frac{Q\left(f_{q_{1}}+f_{q_{2}}\right)\left[\left(\omega
_{0}^{2}-\omega ^{2}\right)^{2}-4\gamma _{0}^{2}\omega
^{2}\right]}{\left[\left(\omega _{0}^{2}-\omega ^{2}\right)^{2}+4\gamma
_{0}^{2}\omega ^{2}\right]^{2}}\Bigg\},
\end{equation}
\begin{equation}
\label{51}
\sigma (\omega )=2A\gamma _{0}\omega
^{2}\Bigg\{\frac{f_{q_{2}}-f_{q_{1}}
}{\left(\omega _{0}^{2}-\omega
^{2}\right)^{2}+4\gamma _{0}^{2}\omega ^{2}}
+\frac{2Q\left(f_{q_{1}}+f_{q_{2}}\right)\left(\omega
_{0}^{2}-\omega ^{2}\right)
}{\left[\left(\omega _{0}^{2}-\omega ^{2}\right)^{2}+4\gamma
_{0}^{2}\omega ^{2}\right]^{2}}\Bigg\},
\end{equation}
but these results are formally the same as the results for the classical
model of the optical activity based on the model of two coupled oscillators
\cite{Vys2}. We assume that the terms, which Janku has
obtained in the quantum mechanical model \cite{Janku} in addition to
the classical model, are only the consequence of used solution method.

The results of the solution of the rotational strengths (\ref{48}) and
(\ref{49}) enable us to express the results for $\bar \rho (\omega )$, $\rho
(\omega )$ and $\sigma (\omega )$, they are given by formulae
(\ref{33}), (\ref{34}) and (\ref{35}), in another form. The formula
for the complex rotatory power can be expressed as
\begin{equation}
\label{52}
\bar \rho (\omega )=\frac{16\pi N\omega ^{2}}{\hbar c}\sum_{\eta
=1}^{2}\sum_{k}\frac{R_{\eta _{0k}}}{\omega _{\eta
_{k0}}^{2}-\omega ^{2}-2i\gamma _{0}\omega }
\end{equation}
and similarly the formulae for $\rho (\omega )$ and $\sigma (\omega )$
are
\begin{equation}
\label{53}
\rho (\omega )=\frac{16\pi N\omega ^{2}}{\hbar c}\sum_{\eta
=1}^{2}\sum_{k}\frac{\left(\omega _{\eta _{k0}}^{2}-\omega ^{2}\right)R_{\eta
_{0k}}}{\left(\omega _{\eta _{k0}}^{2}-\omega ^{2}\right)^{2}+4\gamma
_{0}^{2}\omega ^{2}},
\end{equation}
\begin{equation}
\label{54}
\sigma (\omega )=\frac{32\pi N\omega ^{3}\gamma _{0}}{\hbar c}\sum_{\eta
=1}^{2}\sum_{k}\frac{R_{\eta
_{0k}}}{\left(\omega _{\eta _{k0}}^{2}-\omega ^{2}\right)^{2}+4\gamma
_{0}^{2}\omega ^{2}}.
\end{equation}

We can accept these results as the crystal analogs of the similar
results which we know from the optical activity of molecules
\cite{Moff} or recently of delocalized molecular aggregates \cite{Wag}.
In the future it will be verified
if these results are valid also for other models of coupled
oscillators in the crystal optical activity or by what way they must be
modified.

\label{konec-vys4}


\vspace{1cm}


\begin{thebibliography}{99}
\bibitem{Agr1} Agranovich, V. M.: Opt. i Spektroskop., {\bf 2}, 1957, 738.
\bibitem{Agr2} Agranovich, V. M.: Theory of Excitons (in russian).
Moscow, Nauka 1968.
\bibitem{Tsv} Tsvirko, J. A.: Zhurnal Teor. i Eksp. Fiz., {\bf 38},
1960, 1615.
\bibitem{Kato} Kato, T., Tsujikawa, I., Murao, T.: J. Phys. Soc. Japan, {\bf
14}, 1973, 763.
\bibitem{Chand1} Chandrasekhar, S.: Proc. Ind. Acad. Sci., {\bf A35},
1952, 103.
\bibitem{Chand2} Chandrasekhar, S.: Proc. Roy. Soc, {\bf A 259}, 1961, 531.
\bibitem{Vys1} Vy\v s\'\i n, V.: Proc. Phys. Soc., {\bf 87}, 1966, 55.
\bibitem{Vys2} Vy\v s\'\i n, V.: Opt. Communications, {\bf 1}, 1970, 307.
\bibitem{Janku} Jank\accent23 u, V.: Optica Acta, {\bf 16}, 1969, 225.
\bibitem{Nelson} Nelson, D. F.: JOSA, {\bf B6}, 1989, 1110.
\bibitem{Kuhn} Kuhn, W.: Stereochemie. Leipzig, F. Deutsche 1933.
\bibitem{Davyd} Dawydow, A. S.: Quantenmechanik. Berlin, VEB Deutsche
Verlag der Wissenschaft 1981.
\bibitem{Deutsche} Deutsche, C. W., Lightner, D. A., Woody, W.,
Moscowitz, A.: Ann. Rev. Phys. Chem., {\bf 20}, 1969, 407.
\bibitem{Vysr} Vy\v s\'\i n, V., Vy\v s\'\i n, I.: Acta Univ. Palacki.
Olomuc., Fac. Rer. Nat. (1997), Physica {\bf 36} (in print).
\bibitem{Moff} Moffitt, W., Moscowitz, A.: J. Chem. Phys., {\bf 30},
1959, 648.
\bibitem{Wag} Wagersreiter, T., Mukamel, S.: J. Chem. Phys., {\bf 105},
1996, 7995.






\end{thebibliography}
\end{document}